\documentclass[lengthestimate,tighten,amssymb,prl,aps,twocolumn,byrevtex,floatfix,tightenlines,showpacs]{revtex4}

\usepackage{epsfig}

\newif\ifpdf
\ifx\pdfoutput\undefined
\pdffalse 
\else
\pdfoutput=1 
\pdftrue
\fi

\ifpdf
\usepackage[pdftex]{graphicx}
\else
\usepackage{graphicx}
\fi

\begin{document}
\ifpdf
\DeclareGraphicsExtensions{.pdf, .jpg, .tif}
\else
\DeclareGraphicsExtensions{.eps, .jpg}
\fi

\title{High-Fidelity $Z$-Measurement Error Correction of Optical Qubits}
\author{J. L. O'Brien, G. J. Pryde, A. G. White and T. C. Ralph}
\affiliation{Centre for Quantum Computer Technology, Department of Physics, University of Queensland 4072, Australia}
\date{\today} 

\begin{abstract}
We demonstrate a quantum error correction scheme that protects against accidental measurement, using an encoding where the logical state of a single qubit is encoded into two physical qubits using a non-deterministic photonic CNOT gate. For the single qubit input states $|0\rangle$, $|1\rangle$, $|0\rangle\pm|1\rangle$, and $|0\rangle\pm i|1\rangle$ our encoder produces the appropriate 2-qubit encoded state with an average fidelity of $0.88\pm0.03$ and the single qubit decoded states have an average fidelity of $0.93\pm0.05$ with the original state. We are able to decode the 2-qubit state (up to a bit flip) by performing a measurement on one of the qubits in the logical basis; we find that the 64 1-qubit decoded states arising from 16 real and imaginary single qubit superposition inputs have an average fidelity of $0.96\pm0.03$.
\end{abstract}
\maketitle
One of the greatest promises of quantum information science is the exponential improvement in computational power offered by a quantum computer for certain tasks \cite{nielsen}. Experimental progress has been made in NMR \cite{va-nat-414-883}, ion trap \cite{sc-nat-422-408,le-nat-422-412}, cavity QED \cite{tu-prl-75-4710}, superconducting \cite{ya-nat-425-941}, and spin \cite{cl-ptrsa-361-1451} qubits. A relatively recent proposal by Knill, Laflamme, and Milburn (KLM) is linear optics quantum computing (LOQC) \cite{kn-nat-409-46} in which quantum information is encoded in single photons, and the non-linear interaction required for two photon gates is realised through conditional measurement \cite{pi-pra-68-032316,ob-nat-426-264,ga-quant-ph/0404107,zh-quant-ph/0404129}. A major challenge for all architectures is fault tolerance 
 which will require quantum error correction (QEC) \cite{sh-pra-52-2493,st-prl-77-793}: a logical qubit $|\psi\rangle_L$ is encoded in a number of physical qubits such that joint measurements of the qubits can extract information about errors without destroying the quantum information. A 5-qubit encoding against all 1-qubit errors has been demonstrated in NMR \cite{kn-prl-86-5811} and the 2-qubit encoding $\alpha|00\rangle+\beta|11\rangle$ has been demonstrated with polarisation single photon qubits \cite{pi-pra-69-042306}. 

A simple QEC code is the one introduced by KLM that protects against a computational basis measurement---$Z$-\emph{measurement}---of one of the qubits \cite{kn-nat-409-46}:
\begin{equation}
     |\psi\rangle_L=\alpha |0 \rangle_{L} + \beta|1 \rangle_{L}  = \alpha (|00 \rangle + |11 \rangle) + \beta (|01 \rangle + |10 \rangle)
     \label{eq1}
\end{equation}
This is a parity encoding: $|0\rangle_L$ is represented by all even parity combinations of the 2 qubits; $|1\rangle_L$ by all the odd parity combinations. If a $Z$-measurement is made on either of the physical qubits and the result ``0'' is obtained, then the state collapses to an unencoded qubit, but the superposition is preserved; if the result is ``1'' a bit-flipped version of the unencoded qubit is the result. This generalizes straightforwardly to an $n$-qubit code: if a $Z$-measurement occurs on any of the qubits, the encoding is simply reduced to $n-1$ qubits. This type of QEC is a key tool for scale up of LOQC circuits: KLM showed that their non-deterministic, teleported CNOT gate \cite{kn-nat-409-46}  fails by performing a $Z$-measurement on one of the qubits (this is also true for equivalent gates \cite{pi-pra-64-062311,ko-pra-63-030301}). Thus by using this QEC technique the qubits can be protected against gate failures and so the effective success rate of gate operations can be boosted. A similar principle underlies alternative LOQC schemes \cite{ni-quant-ph/0402005,br-quant-ph/0405157} in which a quantum computation proceeds by generating a highly entangled state of many qubits---a cluster state---and then performing only single qubit measurements in a basis determined from the outcome of previous measurements \cite{ra-prl-86-5188}. Like the $n$-qubit $Z$-error QEC code, these cluster states are robust against accidental $Z$-measurement. The encoding of Eq. 1 also forms the basic element in an encoding which converts photon loss to a bit flip: $|\psi\rangle_{LL}=\alpha|0\rangle_L|0\rangle_L|0\rangle_L+\beta|1\rangle_L|1\rangle_L|1\rangle_L$. For these reasons it is important to show that qubit states can be encoded with high fidelity and recovered with high fidelity after $Z$-measurements.

Here we report an experimental demonstration of QEC using the 2-qubit code of Eq. (1), excluding the bit flip correction. A single qubit prepared in an arbitrary state $|\psi\rangle=\alpha|0\rangle+\beta|1\rangle$ is input into the target mode of a non-deterministic photonic CNOT gate. An ancilla qubit in the real equal superposition $|0\rangle+|1\rangle$ is input into the control. We use quantum state tomography to determine the resulting 2-qubit encoded state generated for the inputs $|\psi\rangle=|0\rangle$, $|1\rangle$, $|0\rangle\pm|1\rangle$ and $|0\rangle\pm i|1\rangle$ and find an average fidelity of $F=0.88\pm0.03$ with $|\psi\rangle_L$. For the same six 1-qubit inputs, the average fidelity of the reconstructed 1-qubit decoded states is $F=0.93\pm0.05$. Finally, we test the decoding (and encoding) by measuring one or other of the qubits in the computational basis for 16 different real and imaginary superposition inputs and performing 1-qubit quantum state tomography on the second qubit. We find that the average fidelity of this second qubit with the state $|\psi\rangle=\alpha|0\rangle+\beta|1\rangle$ or the bit flipped version of it $|\psi'\rangle=\beta|0\rangle+\alpha|1\rangle$ is $F=0.96\pm0.03$. These results demonstrate that high fidelity $Z$-measurement QEC is possible for a simple 2-qubit code.

\begin{figure}[t]
\begin{center}
\includegraphics*[width=6.5cm]{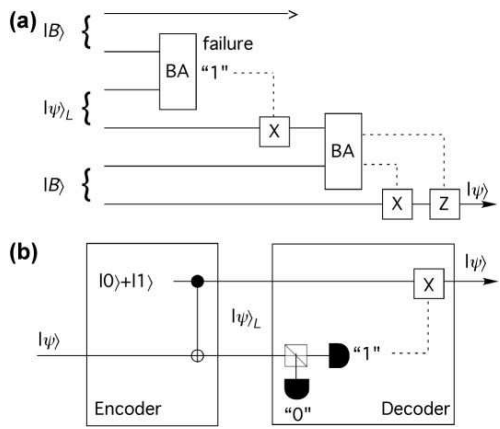}
\caption{Encoding against $Z$-measurement error. (a) A schematic showing how the 2-qubit encoded state $|\psi\rangle_L$ increases the probability of success of the KLM teleportation step ($|B\rangle$=Bell state; BA= Bell analyser). 
(b) A schematic of a circuit for performing the encoding and decoding of the state $|\psi\rangle_L$. The circuit consisits of a CNOT gate with the control input set to $|0\rangle+|1\rangle$ and the arbitrary, 1-qubit state to be encoded $|\psi\rangle$ input into the target. Note that decoding can be achieved by measuring either of the encoded qubits.
}
\label{schematic}
\end{center}
\end{figure}

In the original KLM scheme non-deterministic controlled-phase (CZ) gates are realized through measurement-induced Kerr-like non-linearities, using detection of ancilla modes \cite{kn-nat-409-46}. The scheme of Nielsen also uses CZ gates to build up a cluster state \cite{ni-quant-ph/0402005}. The success probability of CZ gates can be improved with the addition of two 2$n$-photon entangled ancilla states which are used in a scheme to teleport the qubits onto the gate \cite{kn-nat-409-46,go-nat-402-390}. These teleported CZ$_{n^2/(n+1)^2}$ gates work with a probability of $n^2/(n+1)^2$ since the teleportation of each qubit succeeds with probability $n/(n+1)$. When the teleportation steps in a CZ$_{n^2/(n+1)^2}$ fail they cause a $Z$-measurement error. An alternative scheme to increase the success of a CZ$_{n^2/(n+1)^2}$ gate is therefore to encode against $Z$-measurement error. For example, the 2-qubit encoding of Eq. (1)  would allow two attempts at the teleportation, as illustrated in Fig. 1(a), thus doubling the success probability.

Figure 1(a) shows the case were the first Bell analyser (BA) fails and measures the first encoded qubit to be ``1". The logical qubit is then recovered by performing a 1-qubit $X$ gate ($\beta|0\rangle+\alpha|1\rangle\rightarrow\alpha|0\rangle+\beta|1\rangle$). The second BA works, giving the ``11" outcome, and an $X$ gate and a $Z$ gate (($\alpha|0\rangle-\beta|1\rangle\rightarrow\alpha|0\rangle+\beta|1\rangle$) are applied to complete the teleportation. For a LOQC scheme, the 1-qubit $X$ gate required to recover the original state $|\psi\rangle$ from the bit flipped version $|\psi'\rangle$ would require a fast feedforward technique, similar to those demonstrated in Refs. \cite{pi-pra-66-052305,br-quant-ph/0312211}. 

\begin{table}
  \centering 
  \caption{One-qubit input states and the ideal corresponding 2-qubit encoded states.}\label{ }
  \begin{tabular}{|c|c|c|}
\hline
 1 qubit input  & 2 qubit encoded state & Fig. 2  \\
 \hline
 \hline
  $|\psi\rangle=\alpha|0\rangle+\beta|1\rangle$ & $|\Psi\rangle=\alpha|00\rangle+\beta|01\rangle+\beta|10\rangle+\alpha|11\rangle$ &  \\
  \hline
  $|0\rangle$   & $|00\rangle+|11\rangle$ & (a) \\
  $|1\rangle$   & $|01\rangle+|10\rangle$ & (d)\\
  $|0\rangle+|1\rangle$  & $|00\rangle+|01\rangle+|10\rangle+|11\rangle$ & (b)\\
  $|0\rangle-|1\rangle$  & $|00\rangle-|01\rangle-|10\rangle+|11\rangle$ & (e)\\
  $|0\rangle+ i|1\rangle$  & $|00\rangle+i|01\rangle+i|10\rangle+|11\rangle$ & (c)\\
  $|0\rangle- i|1\rangle$  & $|00\rangle-i|01\rangle-i|10\rangle+|11\rangle$ & (f)\\
\hline
\end{tabular}
\end{table}

\begin{figure}[t]
\begin{center}
\includegraphics*[width=8cm]{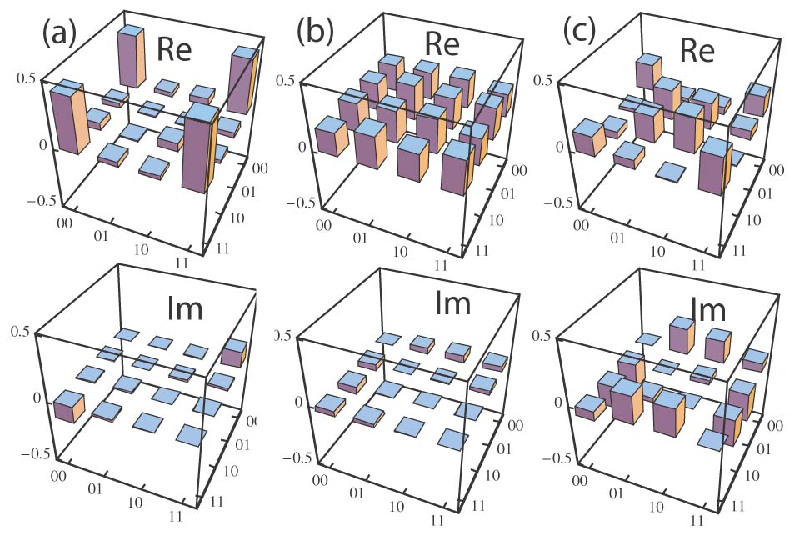}
\includegraphics*[width=8cm]{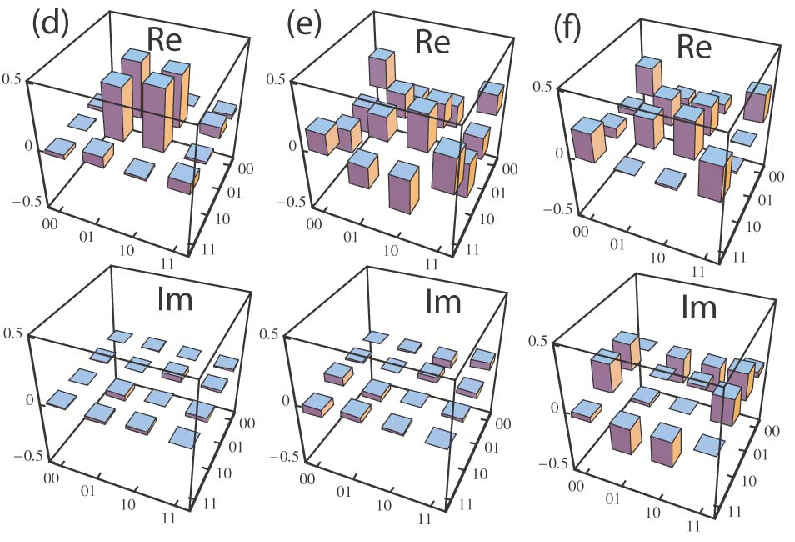}
\caption{Two-qubit encoded states. The real and imaginary parts of the density matrices are shown for the encoded states produced for the 1-qubit inputs given in Table I.}
\label{2qubit}
\end{center}
\end{figure}

Figure 1(b) shows schematically how the encoding and decoding was performed, using a high-fidelity photonic CNOT gate \cite{ob-nat-426-264} and single photon measurement respectively. The CNOT gate operates non-deterministically and successful events are post-selected by coincidence measurements. Pairs of photons were generated through spontaneous parametric down-conversion \cite{enc-note-1} and sent through a polarising beam splitter (PBS) to prepare a highly pure horizontal polarisation state. Qubits are stored in the polarisation state of these two photons using the assignment horizontal $|H\rangle\equiv|0\rangle$ and vertical $|V\rangle\equiv|1\rangle$. The control input is prepared in the equal real superposition $|0\rangle+|1\rangle$ using a half wave plate (HWP) with its optic axis at 22.5$^{\circ}$. An arbitrary $|\psi\rangle$ was prepared using a HWP and quarter wave plate (QWP). The output of the circuit, nominally $|\Psi\rangle_L$, was analysed using standard 2-qubit quantum state tomography \cite{ja-pra-64-052312}. The required 2-qubit measurements were performed using a pair of analysers each consisting of a QWP and HWP followed by a PBS and single photon counting module.

\begin{figure}[t]
\begin{center}
\vspace{0.1cm}
\includegraphics*[width=8.4cm]{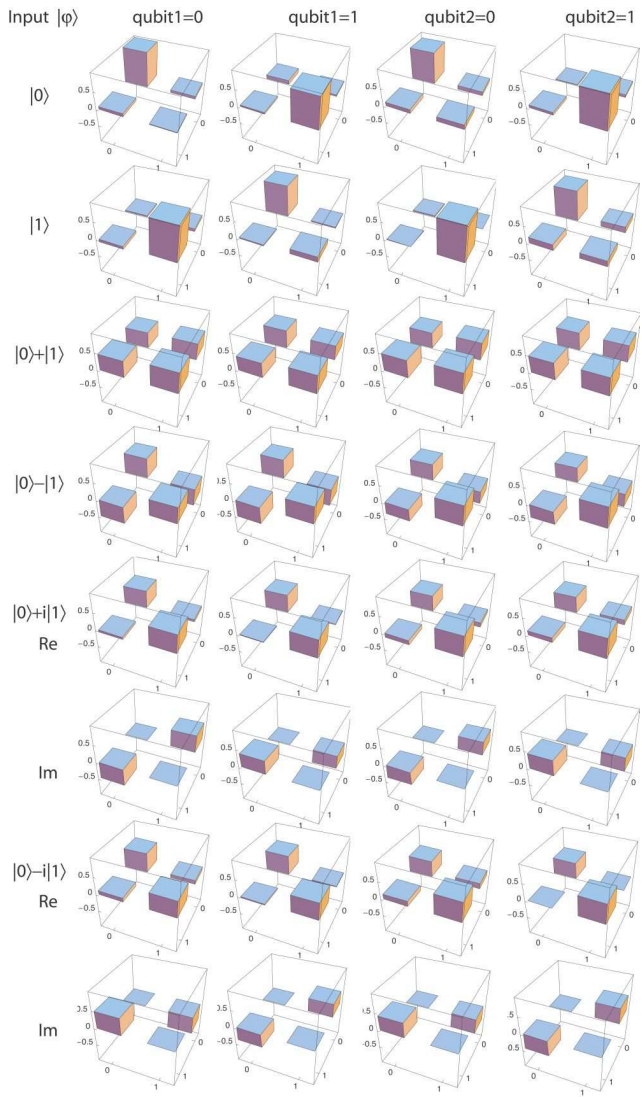}
\caption{One-qubit decoded states. A table of density matrices for the six inputs and four decoding measurements indicated in the figure. The imaginary parts are shown for the $|0\rangle+i|1\rangle$ and $|0\rangle-i|1\rangle$ inputs only.}
\label{1qubit}
\end{center}
\end{figure}

\begin{figure}[t]
\vspace{0.2cm}
\includegraphics*[width=8.3cm]{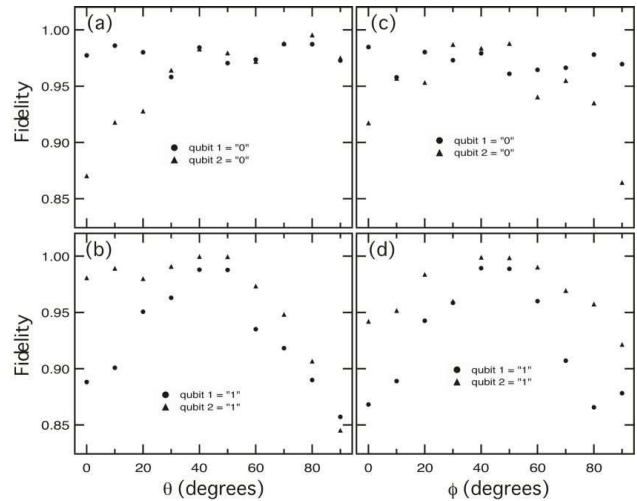}
\caption{One-qubit decoded state fidelities for the 1-qubit input states $cos(\theta)|0\rangle+sin(\theta)|1\rangle$ and $|0\rangle+e^{i(90^{\circ}-2\phi)}|1\rangle$.}
\label{fidelities}
\end{figure}

We firstly confirm that this circuit produces the correct 2-qubit encoded state: Table I shows six 1-qubit inputs (the two eigenstates and four equal superpositions) and the corresponding ideal encoded states. For these six 1-qubit input states we used our encoding circuit to produce a 2-qubit encoded state and performed 2-qubit quantum state tomography on the output. The real and imaginary parts of the density matrices for the six 2-qubit encoded states are shown in Fig. 2. These experimentally measured encoded states have an average fidelity of $F=0.88\pm0.03$ with $|\psi\rangle_L$, demonstrating the high fidelity of the encoding operation.

Next we test how well the encoding has worked by assuming the capability to perform perfect decoding and so theoretically extract the 1-qubit decoded states from the 2-qubit encoded states shown in Fig. 2. The 1-qubit decoded state can be produced in four ways: by measuring either the first or second qubit in the computational basis, and getting the result ``0" or ``1". In Fig. 3 we show all four reconstructed 1-qubit density matrices extracted from each of the six encoded states in Fig. 2. We show the imaginary components only for the imaginary superpositions since in all other cases these components should be zero, and are measured to be relatively small (the average of imaginary components is $0.04\pm0.04$). The average fidelity of these 1-qubit states with $|\psi\rangle$ or $|\psi'\rangle$ is $F=0.93\pm0.05$. Since these states are reconstructed from the six 2-qubit states, this result demonstrates that given perfect decoding, the 1-qubit decoded states are more robust to imperfections in the encoder than the 2-qubit encoded states.

Finally we test the decoding circuit (as well as the encoding circuit) by preparing the 1-qubit input in the unequal amplitude, real superpositions $cos(\theta)|0\rangle+sin(\theta)|1\rangle$, and equal amplitude, variable phase superpositions $|0\rangle+e^{i(90^{\circ}-2\phi)}|1\rangle$, for ${\theta,\phi}$ = {10$^{\circ}$, 20$^{\circ}$, 30$^{\circ}$, 40$^{\circ}$, 50$^{\circ}$, 60$^{\circ}$, 70$^{\circ}$, and 80$^{\circ}$}. Together with the results of Fig. 3 (giving $\theta,\phi$ = 0$^{\circ}$ and 90$^{\circ}$), these states map out two orthogonal great semicircles on the Bloch sphere. For each of these inputs we reconstructed the 1-qubit density matrix \emph{directly} for both measurement outcomes on both qubits, ie four 1-qubit density matrices for each input state. From these density matrices we calculated the fidelity with the ideal state $|\psi\rangle$ or $|\psi'\rangle$. The results are shown in Fig. 4.  The average fidelity for all of these states (excluding $\theta,\phi$ = 0$^{\circ}$ and 90$^{\circ}$) is $0.96\pm0.03$, which is higher than the average for the reconstructed states shown in Fig. 3. This may suggest that errors in the maximum likelihood reconstruction of the 2-qubit states causes the 1-qubit states reconstructed from them to have lower fidelities. The most reliable measure of the fidelity of the QEC would then be the the direct measurement of the 1-qubit decoded states.

The behaviour observed in Fig. 4 can be explained in terms of the classical and non-classical interference requirements of the CNOT gate: the gate splits and recombines the control and target polarisation modes requiring two classical interferences; and non-classically interferes the control $|V\rangle$ and target $|H\rangle-|V\rangle$ modes \cite{ob-nat-426-264}. In practice, the visibility of each of these interferences is sub-unity and thus contributes to errors in the operation of the gate: The encoder works well for $\theta=\phi=45^{\circ}$, i.e., the input state $|0\rangle+|1\rangle$ ($=|H\rangle+|V\rangle$) since no non-classical interference is required, and only one classical interference is required for the control. The encoder also works well when qubit 1 (the output of the control mode) is measured to be $|0\rangle$ since, again, only one classical interference is required. Finally, the encoder works well for $\theta$ close to zero, which is expected from Fig. 1(a) which shows that the $|01\rangle\langle01|$ population, which is the main contributor to errors in this case, is very small.

This 2-qubit $Z$-error encoding could be extended to an $n$-qubit encoding using additional CNOT gates with single photons prepared in the $|0\rangle+|1\rangle$ state as the control input, and any of the already encoded qubits as the target. A very similar technique can be used to build up a cluster state using CZ gates \cite{ni-quant-ph/0402005}. Note, however, that our CNOT gate succeeds with probability 1/9 and is destructive \cite{ra-pra-65-062324}. In order to build up the $Z$-error encoding or a cluster state in a scalable way KLM-type level 1 or level 2 ancilla-assisted teleportation to make a CZ$_{1/4}$ or CZ$_{4/9}$, respectively, or similar techniques \cite{br-quant-ph/0405157} would be needed.

In conclusion, we have experimentally demonstrated a high fidelity realisation of a 2-qubit $Z$-measurement QEC scheme (up to implementation of  an $X$ gate). For a representative set of input states the average fidelity of the 1-qubit decoded state is $0.96\pm0.03$. Our scheme uses a non-deterministic CNOT gate operating on the polarisation states of single photon qubits, and is therefore a non-deterministic encoding and will not be useful in its own right for a scalable quantum computer. However, this does provide a proof of principle of an encoding against a realistic error in linear optics quantum computing. The technique can be extended to a larger encoded state or to the creation of a cluster state by using CZ gates.
\vspace{0.3cm}

This work was supported by the Australian Research Council, the US National Security Agency and Advanced Research and Development Activity under Army Research Office contract number DAAD 19-01-1-0651.


\begin{thebibliography}{27}
\bibitem[{enc()}]{enc-note-1}
\bibinfo{note}{Pairs of energy degenerate non-entangled photons of wavelength
  $\lambda=702.2$ nm were generated in a non-linear $\beta$-barium-borate (BBO)
  crystal through spontaneous parametric down conversion of a $\lambda=351.1$
  nm, $P\approx900$ mW pump beam using the setup described in
  \cite{ob-nat-426-264}. A coincidence window of 1 ns was used and no
  correction for accidental counts was made.}
\end{thebibliography}
\end{document}